\documentclass[lettersize,journal]{IEEEtran}
\usepackage{amsmath,amsfonts}
\usepackage{algorithmicx}
\usepackage{algorithm}
\usepackage{algpseudocode}
\usepackage{array}
\usepackage[caption=false,font=normalsize,labelfont=sf,textfont=sf]{subfig}
\usepackage{textcomp}
\usepackage{stfloats}
\usepackage{url}
\usepackage{verbatim}
\usepackage{graphicx}
\usepackage{cite}
\usepackage{dcolumn}
\usepackage{bm}
\usepackage{xcolor}
\usepackage{nicematrix}
\usepackage{braket}
\usepackage[normalem]{ulem}
\usepackage{comment}
\usepackage[export]{adjustbox}
\usepackage{booktabs}
\usepackage{soul}
\sethlcolor{lime}

\definecolor{Gray}{gray}{0.9}

\begin{document}

\title{Q-fid: Quantum Circuit Fidelity Improvement with  LSTM Networks}

\author{Yikai~Mao,
        Shaswot~Shresthamali,
        and~Masaaki~Kondo
\thanks{Y. Mao, and M. Kondo are with the Graduate School of Science and Technology, Keio University, Yokohama, Kanagawa 223-8522, Japan.  (e-mail: ykmao@acsl.ics.keio.ac.jp)}
\thanks{S. Shresthamali is with the Graduate School of Information Science and Electrical Engineering, Kyushu University, Nishi-ku, Fukuoka 819-0395, Japan. He is also a visiting researcher in Kondo Laboratory, Keio Univerisity. (e-mail: shaswot@acsl.ics.keio.ac.jp)}
\thanks{M. Kondo is also with RIKEN Center for Computational Science, Kobe, Hyogo 650-0047, Japan. (e-mail: kondo@acsl.ics.keio.ac.jp)}
}



\maketitle

\begin{abstract}
The fidelity of quantum circuits is influenced by several factors, including hardware characteristics, calibration status, and the transpilation process, all of which impact their susceptibility to noise. However, existing methods struggle to estimate and compare the noise performance of different circuit layouts due to fluctuating error rates and the absence of a standardized fidelity metric. In this work, we introduce Q-fid, a Long Short-Term Memory (LSTM) based fidelity prediction system accompanied by a novel metric designed to quantify the fidelity of quantum circuits. Q-fid provides an intuitive way to predict the noise performance of NISQ quantum circuits. Our approach frames fidelity prediction as a Time Series Forecasting problem to analyze the tokenized circuits, capturing the causal dependence of the gate sequences and their impact on overall fidelity. Additionally, the model is capable of dynamically adapting to changes in hardware characteristics, ensuring accurate fidelity predictions under varying conditions. Q-fid achieves a high prediction accuracy with an average RMSE of \(0.0515\), up to \(24.7\times \) more accurate than the Qiskit transpile tool \texttt{mapomatic}. By offering a reliable method for fidelity prediction, Q-fid empowers developers to optimize transpilation strategies, leading to more efficient and noise-resilient quantum circuit implementations.
\end{abstract}



\section{Introduction}
%
%
%
%

\IEEEPARstart{B}{y} using quantum computers that exploit the principles and phenomena of quantum mechanics, it may be possible to achieve superpolynomial or even exponential speedups for traditionally hard computing problems \cite{simon}. With quantum computers, in addition to representing the classical bits \(0/1\) using its computational basis \(\ket{0}/\ket{1}\), the qubits also have the ability to switch to any desired basis, elevating the calculation to a higher dimension. Moreover, by entangling multiple qubits, the computation can be performed simultaneously in an exponentially larger space.

However, present-day quantum computers have a limited amount of qubits with limited interconnectivity and very short coherence lifetimes (in order of milliseconds) \cite{qubit_life_0, qubit_life_1}. Furthermore, the gate operations and readout are very susceptible to external noise. To overcome the noise limitations of today's machines, Noisy Intermediate-Scale Quantum (NISQ) \cite{nisq} offers a viable solution by using error-mitigation techniques to compensate for fragile qubits \cite{error_mitigation, ibm_mitigation, qulatis, nisq_techs, qecool}.

To implement a quantum circuit on a real NISQ processor, the circuit must be mapped onto the available physical qubits according to the hardware's connectivity map, a process called transpilation \cite{allocation, not_equal}. There are several challenges during this process: First, the qubits in one processor are not identical. Each of them has unique physical properties that define their noise characteristics, and they can change depending on the time of operation. This causes the quantum circuits to exhibit different noise performances when they are placed on different qubits. Second, due to the limited qubit connectivity, some quantum circuits need to be modified from the original design before implementing on real hardware. For example, adding SWAP gates to bring two physically distant qubits together for a CNOT operation. Therefore, the accuracy of a quantum circuit can get compromised even if it is placed on high-quality qubits, due to the additional gate noise introduced by circuit transpilation.


To overcome these challenges, we need a tool to accurately pick a high-fidelity circuit-to-qubit layout from all the transpilation options. This way we can efficiently retrieve high-quality measurements, without wasting quantum computing resources on testing error-prone layouts. Additionally, a promising direction for implementing large quantum circuits on NISQ processors is by breaking them down into smaller circuits \cite{nisq_future, knitting}. In this case, picking a higher-fidelity layout for the smaller circuit is even more important because it will eventually affect the final output for the full circuit. 

One such tool to find a high-fidelity layout is \texttt{mapomatic}, built within the Qiskit transpiler \cite{mm}. \texttt{mapomatic} estimates fidelity by accumulating the individual gate error rates, obtained from the Randomized Benchmarking (RB) experiments \cite{rb_0}. This approach has some severe limitations. For example, the RB experiments must be performed frequently to keep the error rates up to date. Also, the noise model of a full quantum circuit is more complex than the accumulation of the individual error rates \cite{rb_bad}. Due to these limitations, \texttt{mapomatic} cannot estimate the fidelity accurately and it only gives the relative performance comparison for a set of transpilations.

\begin{figure}[tb]
    \centering
    \includegraphics[width=\columnwidth]{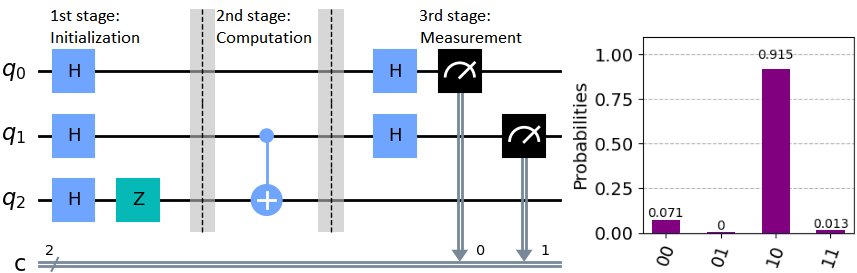}
    \caption{Bernstein-Vazirani Algorithm. The expected output is \(q_1q_0=\) $\ket{10}$ (\(q_2\) is not measured). Output distribution after running the circuit for 1,024 times on real hardware (\textit{ibm\textunderscore nairobi}) is plotted on the right, the correct state $\ket{10}$ has been measured 937 times, but the wrong measurements $\ket{00}$ and $\ket{11}$ also appears in the distribution due to noise.}
    \label{1a_BV}
\end{figure}

In this work, we present a practical metric: d-$R^2$, that intuitively represents the fidelity of the output distribution from a quantum circuit. Here, fidelity refers to the computational accuracy of a quantum circuit/gate (a measurement of how closely the actual output of the quantum circuit matches the expected output), which is different from the qubit state fidelity (a quantification of the overlap between two states). Using this metric, we then develop an LSTM-based system: Q-fid, to accurately predict the fidelity of a quantum circuit. It eliminates the need for frequent RB experiments by actively learning qubit/gate operations from historical circuit execution data, without any separate input of system calibration parameters or error rates. Traditionally, a NISQ circuit requires hundreds to thousands of shots on a real processor to statistically estimate its correct output. With Q-fid, we can choose to only execute circuits that have higher fidelity and obtain the solution with fewer shots, thus saving precious quantum computing resources.

\begin{figure*}[t]
    \centering
    \includegraphics[width=\textwidth]{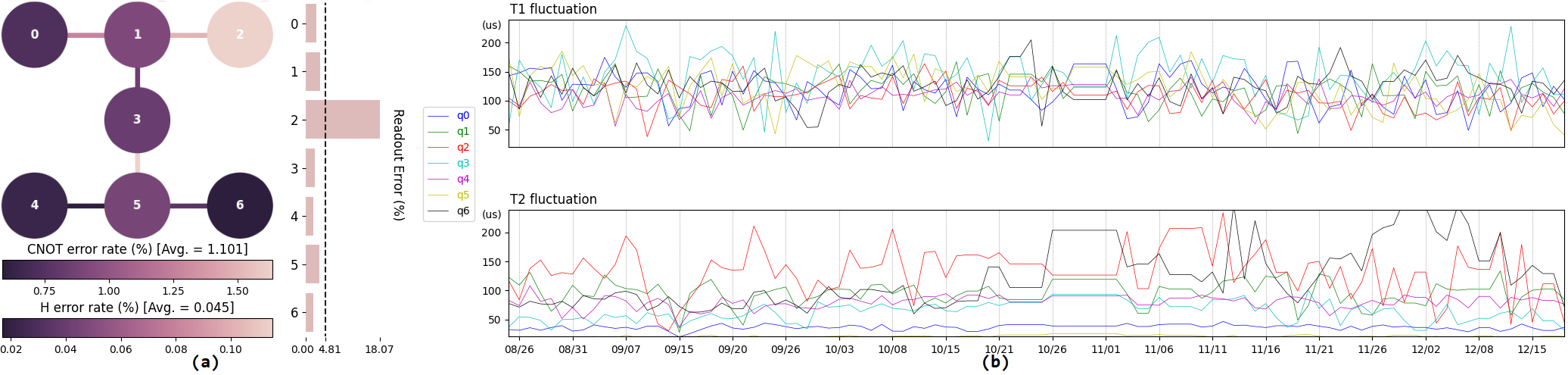}
    \caption{(a) Error map of \textit{ibm\textunderscore nairobi} on Dec. 9, 2022, generated by the IBM Q platform \cite{ibmq}. (b) \(T_1/T_2\) fluctuation of the 7 physical qubits inside \textit{ibm\textunderscore nairobi}. 100 data points were collected from Aug. 25 to Dec. 18, 2022. Note that data is not continuous due to scheduled/unscheduled system maintenance, for example around Nov. 1.}
    \label{2b_map_t1t2}
\end{figure*}

The contributions of this work are listed below:
\begin{itemize}
\item We propose a simple and intuitive method to model a quantum circuit using text. This method can be applied to any gate-based quantum processor and enables feeding the quantum circuits directly into an LSTM neural network.
\item We present the discrete coefficient of determination (d-$R^2$) to evaluate the noisy output distribution of a quantum circuit. d-$R^2$ uses the uniform distribution as the worst-case output, offering a reasonable baseline for comparing NISQ algorithm fidelities. 
\item We show that LSTM is effective in learning the error properties of a qubit and a quantum gate. The trained system, Q-fid, can predict the performance of a quantum circuit without any separate input of hardware calibration data or gate error rates.
\item We provide a framework to use LSTM for on-the-fly quantum circuit fidelity estimation, including the architecture of the neural network, how to build the dataset using Randomized Benchmarking, and the training workflow.
\item Experiments using real NISQ algorithms show that because Q-fid can accurately predict the d-$R^2$ score of quantum circuits, we can retrieve more high-fidelity, usable transpilations than \texttt{mapomatic} from a large set of transpiled circuits.
\end{itemize}



\section{Background}

\subsection{NISQ Circuits and Processors\label{qc}}

In a gate-based quantum computer, the quantum bits (qubits) are manipulated by a sequence of quantum gates as described by the quantum circuit (QC) \cite{mandi}. These gates change the state of the qubit and the transformations can be expressed by unitary matrices i.e., the computation is reversible. Fig. \ref{1a_BV} shows a sample implementation of the Bernstein-Vazirani Algorithm \cite{bv} divided into three stages. 1. State preparation: The qubits are initialized into the superposition state $\ket{-++}$ from $\ket{000}$. 2. Computation: Once initialized, the quantum computation (CNOT as in this example) is performed. 3. Measurements: The qubits are returned to their original basis and measured into the classical registers \(c_1c_0\) to store the output.

NISQ processors have limited qubit availability, both in terms of qubit quantity and quality \cite{nisq}. Fig. \ref{2b_map_t1t2}(a) gives the architecture of the 7-qubit quantum processor \textit{ibm\textunderscore nairobi} along with three major forms of possible error: readout, single-qubit error, and CNOT error. The qubit quality is commonly characterized by their \(T_1/T_2\) constants. \(T_1\) is called the coherence time and \(T_2\) is the decay time, which describes how long a qubit relaxes to the ground state and how long it can hold its phase. However, as shown in Fig. \ref{2b_map_t1t2}(b), because qubits are very sensitive to multiple sources of noise, their \(T_1/T_2\) constants can fluctuate significantly, and it is hard to predict which physical qubit is more stable than others at a given time.

\subsection{Circuit Transpilation}

Due to the limited connectivity of current NISQ processors, it is not always possible to map a QC onto the processor directly \cite{1604.01401}. For example, the physical CNOT links exist only between the neighboring qubits in Fig. \ref{2b_map_t1t2}(a). So if a QC requires a CNOT gate between qubit 0 and qubit 3, additional operations must be added to the QC to compensate for the missed connection. This process of translating a hardware-agnostic QC description to implement in a given hardware platform is referred to as transpilation.

To perform a long-distance CNOT gate in a superconducting quantum computer like \textit{ibm\textunderscore nairobi}, the transpiler can insert SWAP gates to switch the position of the nonadjacent physical qubits to use the existing CNOT links \cite{2212.05666}. However, in addition to connectivity restrictions, real quantum processors also have limited single-qubit gate availability. For example, \textit{ibm\textunderscore nairobi} only supports four single-qubit gates: ID, RZ, SX, X. Therefore, To perform a single-qubit operation like the Hadamard gate, the transpiler will also need to decompose the operation into the gates available in the processor.

\begin{figure}[tb]
    \centering
    \includegraphics[width=\columnwidth]{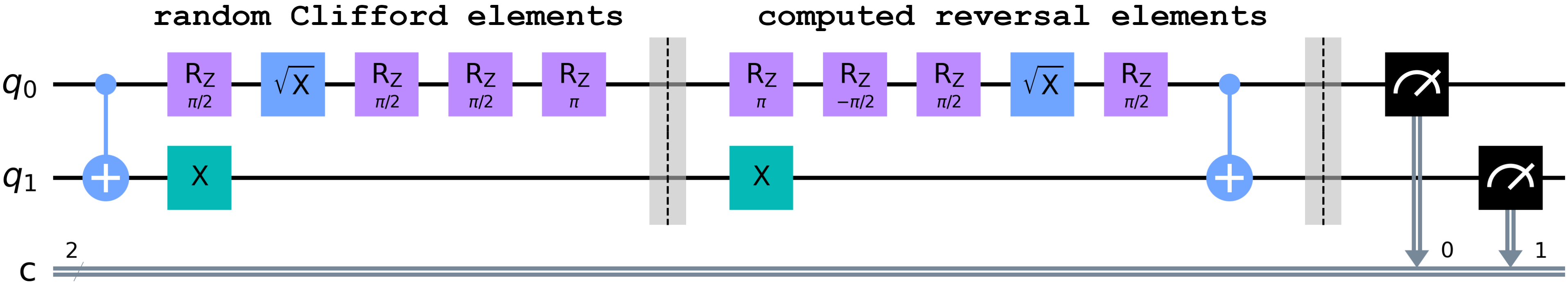}
    \caption{Example of a Randomized Benchmarking circuit. The first part is the randomly generated Clifford gates, followed by the calculated reversal gates. The final measurement should be \(\ket{00}\) if no errors occur.}
    \label{3c_rb}
\end{figure}

\subsection{Randomized Benchmarking}

Randomized Benchmarking (RB) \cite{rb_0, rb_1} is an experiment to estimate the error rates of the set of common quantum gates, usually called the Clifford gates. For IBM Q devices, this Clifford gate group includes [X, Z, P, H, CNOT, CZ, SWAP] \cite{clifford}. Based on the reversible principle of quantum gates, the RB experiment first generates a QC containing random quantum gates in the Clifford group, then it calculates a complementary gate sequence that can reverse the computation performed in the first QC. An example RB circuit is given in Fig. \ref{3c_rb}. By applying the two generated QCs back-to-back, the full circuit is equivalent to an identity operator so ideally any qubits involved in this circuit should never change their states when measured at the end. 

However, when executing an RB circuit on a NISQ processor, it is possible that the qubits cannot return to their initial state due to the noisy nature of the hardware. Therefore, by repeatedly running RB circuits and measuring the qubit outcomes, we can estimate the average fidelity of the processor, and use that information in turn to predict the gate error rate, either for 2-qubit gates or single-qubit gates.

\subsection{Long Short-term Memory Network}

Neural Networks (NN) have demonstrated state-of-the-art performance in various tasks including Computer Vision (CV) and Natural Language Processing (NLP). Among numerous NN architectures, Long Short-term Memory (LSTM) networks \cite{lstm} have been a popular choice for tasks related to time series processing, for example, weather forecasting and sentiment prediction. 

We leverage the ability of LSTMs to learn temporal relationships in sequential data to estimate circuit fidelity. It is very easy to see that quantum circuits are essentially sequential gate operations applied on qubit(s). There is an inherent temporal sequence in the execution of the circuit on qubits. Furthermore, the fidelity of the entire circuit not only depends on the individual gate/qubit characteristics but also on how they interact with each other as the circuit progresses towards completion, much like words in a sentence or notes of a musical score. Since LSTMs are trained to recognize these temporal relationships, we use them to estimate circuit fidelity in our work. To our knowledge, this is the first work that uses LSTM for fidelity estimation.

\begin{figure}[tb]
    \centering
    \includegraphics[width=\columnwidth]{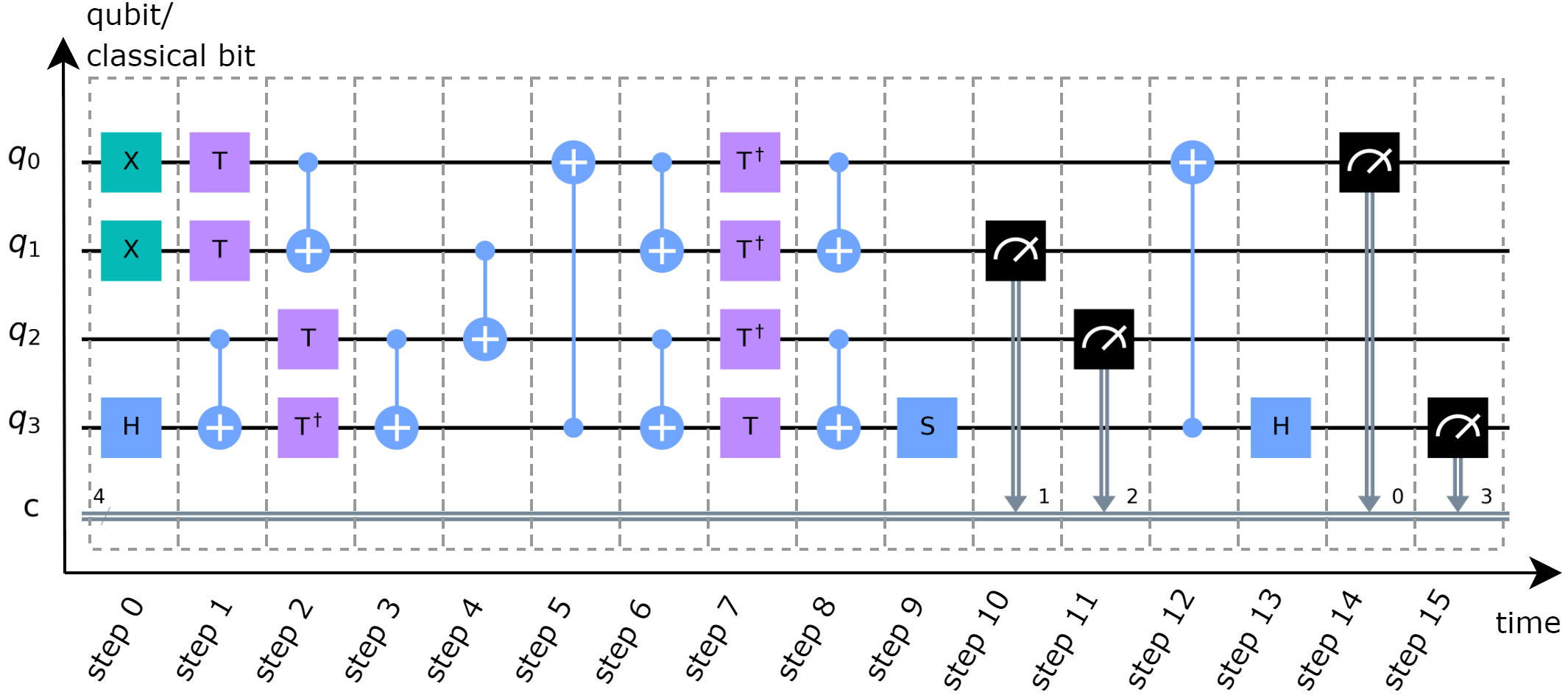}
    \caption{Quantum circuit placed in a coordinate system with the y-axis as qubit/classical bit, and the x-axis as time. It can be viewed as the qubits being manipulated through each timestep from the left to the right on the time axis.}
    \label{2d_timeseries}
\end{figure}

\begin{figure*}[tbp]
    \centering
    \includegraphics[width=\textwidth]{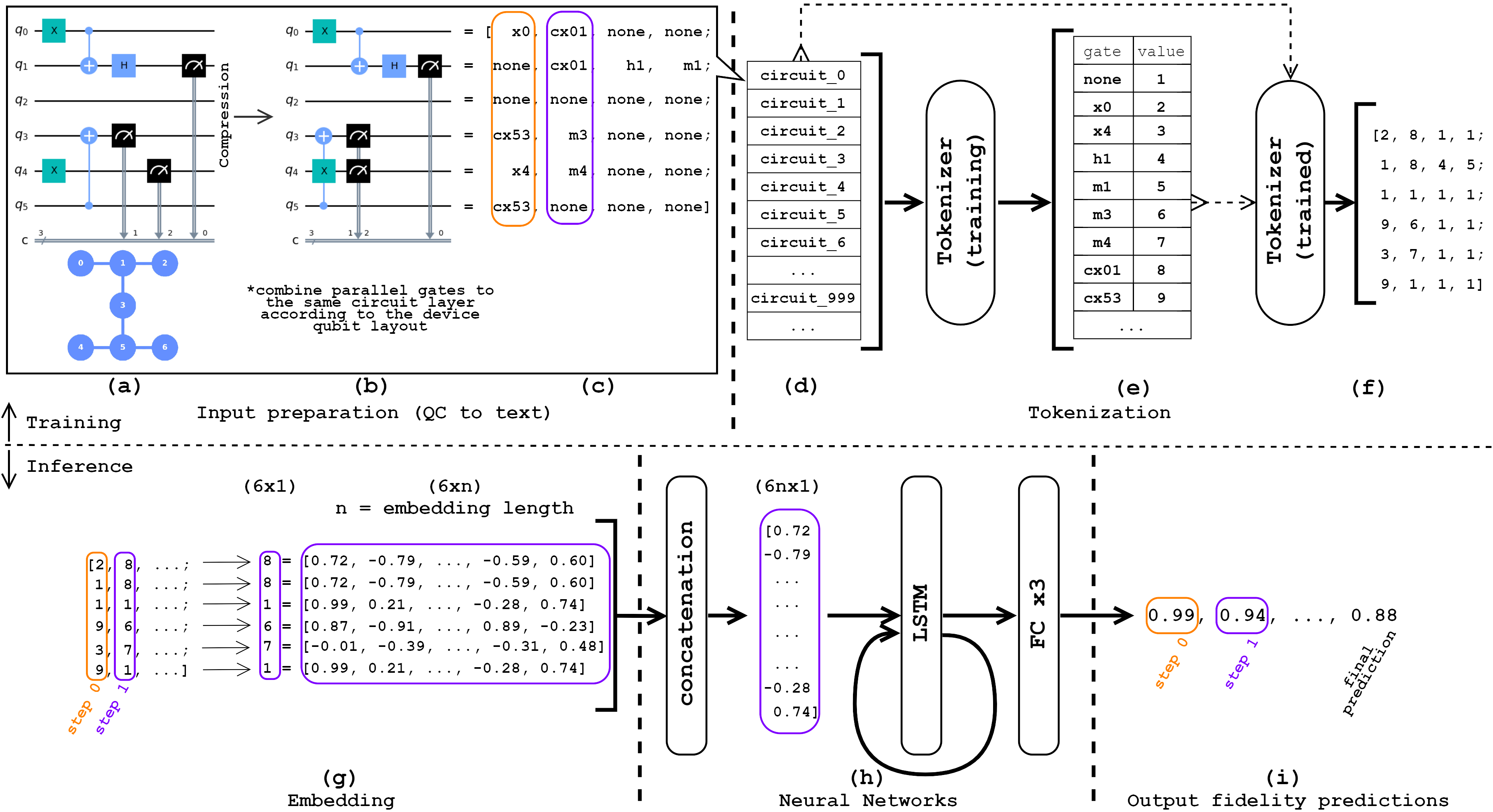}
    \caption{Overview of Q-fid's workflow. The first stage is input preparation (a, b, c), where the QCs are compressed and mapped to their corresponding text label. The compression causes the gates to stack visually but they are executed individually on hardware because the qubits are not physically connected (as shown in Fig. S4(a)). The second stage is tokenizer training (d, e), the tokenizer will look through all the QCs in the dataset and assign each text label with an integer according to their appearing frequency. After the tokenizer is trained, it will use its internal dictionary to translate all of the QCs in the dataset to an integer representation, which is called tokenization (f). Q-fid's neural network uses embedding (g) and LSTM (h) layers to encode the integer labels and then extract sequential and relationship information from the QC, each prediction (i) is associated with one timestep in the QC. After the final timestep, the output will be the fidelity prediction of the full QC, represented as a d-$R^2$ score.}
    \label{3a_flow}
\end{figure*}

As Fig. \ref{2d_timeseries} shows, the x-axis can be used to indicate the timesteps of a QC from start to end, with each timestep modeled as a ``QC layer'' containing all the gates in the y-axis. The layers act on the same set of qubits and have fixed widths, so the QC can be described as a two-dimensional time series with its width equal to the number of qubits in the circuit, and its length equal to the number of layers (sometimes referred to as the depth) of the circuit.

\section{Proposed Q-fid Framework}

\subsection{Framework Description}

Q-fid is an LSTM network that takes a QC as input and predicts the fidelity of its output distribution. Inspired by one of the most popular LSTM applications, Sentiment Analysis \cite{lstm_sentiment}, the workflow of Q-fid is very similar to the workflow of many common NLP tasks. In these tasks, the LSTM takes a sentence as input and gives a prediction based on different objectives and contexts. For example, translate the sentence or guess the sentiment. Although every individual word has its own definition, once combined together, they become elements of a single sentence where the meaning of the sentence must be inferred from the relationship between all of the words, i.e., the contexts.

This concept fits surprisingly well when applied to a QC: Although every individual quantum gate has its own operation and noise characteristics, once combined together, they become elements of a single circuit where the computation and fidelity of the qubits must be measured after all quantum gates have been executed. Similar to a sentence where different orders of the same words can express different sentiments, the same set of quantum gates can perform different computations and express different fidelity depending on their order in a circuit. Q-fid uses LSTM to catch this long-term temporal dependence and noise characteristics inside a QC.

At the core of the proposed Q-fid framework is an LSTM network with a lightweight architecture. The input layers first perform general pre-processing of the input QCs, then they are passed into the LSTM layer to extract long-term and short-term noise dependencies between the gates inside the QC. Finally, the output from the LSTM layer is passed into a series of Fully-Connected layers to generate the final prediction of the circuit fidelity. We use a data-driven approach to train the Q-fid system, and the noise characteristics of the hardware are approximated by the 700,000+ parameters inside the LSTM network. The number of input neurons is decided by the number of qubits of the quantum processor. In other words, a larger quantum processor will have a larger Q-fid neural network with more inputs attached to it. An overview of Q-fid's LSTM architecture is shown in Fig. \ref{3a_flow} from (g) to (I). 

A distinctive feature of Q-fid is that it does not require any explicit input of the processor's calibration data (\(T1/T2\) frequency, etc.) to make predictions, since the LSTM infers the noise characteristics from the input QC during training. This hardware-agnostic feature gives Q-fid several advantages over the other calibration-based prediction systems. First, because the hardware descriptions are abstracted away, Q-fid can work with any gate-model quantum computer regardless of whether the qubits are superconducting or trapped ion. Second, since the user does not need to specify any device calibration data, Q-fid can be trained dynamically during regular workload and adapts to the ever-changing device characteristics, all without interruptions caused by calibration or maintenance jobs.

\begin{figure*}[tb]
    \centering
    \includegraphics[width=\textwidth]{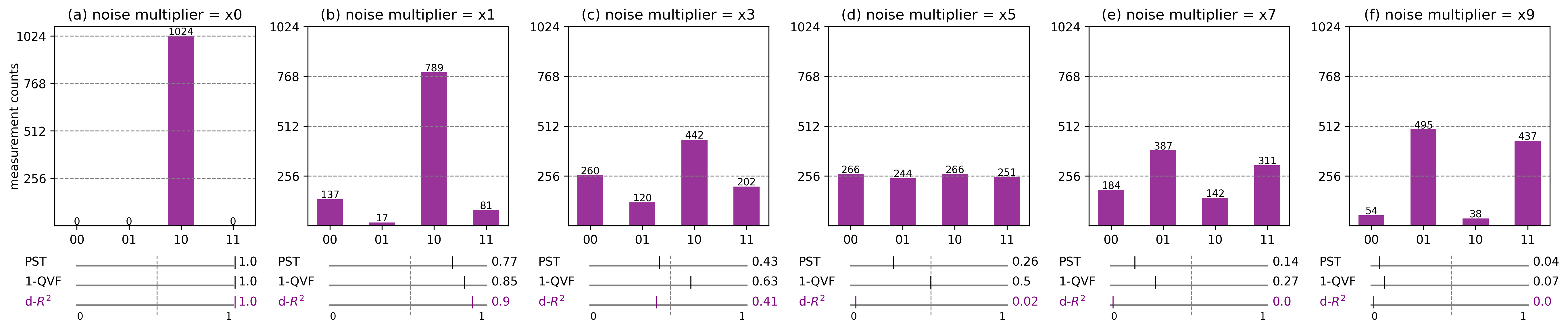}
    \caption{Output distributions after executing the QC in Fig. \ref{1a_BV} with 1,024 shots using the Qiskit Aer simulator with varying noise intensity. The base error rates for gates, measurements, and reset are set as $0.05n$, $0.1n$, and $0.03n$. $n$ is the noise multiplier where \(\times0\) means noise-free and \(\times10\) is the maximum possible value. The comparison of three QC fidelity metrics is shown below the bar graph, where 1 means perfect circuit fidelity. (a) is the noise-free output, the correct output state should be $\ket{10}$. (b) to (f) are the outputs under different noise multiplier values. The output in (d) is a uniform distribution, which gives zero information on which state might be the correct state. (f) is considered as a faulty output, the circuit is not doing the intended quantum operations so the wrong states are appearing more than the correct state.}
    \label{3d_fids}
\end{figure*}

\subsection{Discrete Coefficient of Determination (d-$R^2$)}\label{dr2}

The LSTM updates its parameters by comparing the differences between the true output probability distribution of the QC and the observed output distribution. We thus need a single metric that represents this difference so that we can feed it into the loss function of the LSTM network. In this paper, we apply a modified version of the Coefficient of Determination ($R^2$) as our metric for evaluating noisy QC fidelity. $R^2$ is commonly used in regression analysis to show the goodness of fit \cite{r2_good}, where $R^2=1$ indicates a perfect fit and $R^2=0$ indicates that the fitted line does not represent the original data at all.

Recent works \cite{not_equal, pst_modeling} use the Probability of Successful Trials (PST) as a metric to quantify the performance of a noisy QC, which is defined as the ratio of the number of successful trials to the total number of trials. Although PST is easy to calculate, it does not give the user enough information to analyze the circuit. For example, Fig. \ref{3d_fids}(e) shows a distribution where PST gives a fidelity score of 0.14. However, the user cannot distinguish whether this low fidelity is caused by $\ket{00}$ or $\ket{10}$, due to their similar measurement counts. Other metrics have been proposed to replace PST, for example, the Quantum Vulnerability Factor (QVF) \cite{qvf} and the Total Variation Distance (TVD) \cite{charter}. However, they do not offer a clear definition when multiple correct states are expected, which often happens for common NISQ algorithms like QAOA \cite{qaoa} and VQE \cite{vqe}.

\begin{figure}[tb]
    \centering
    \includegraphics[width=1\columnwidth]{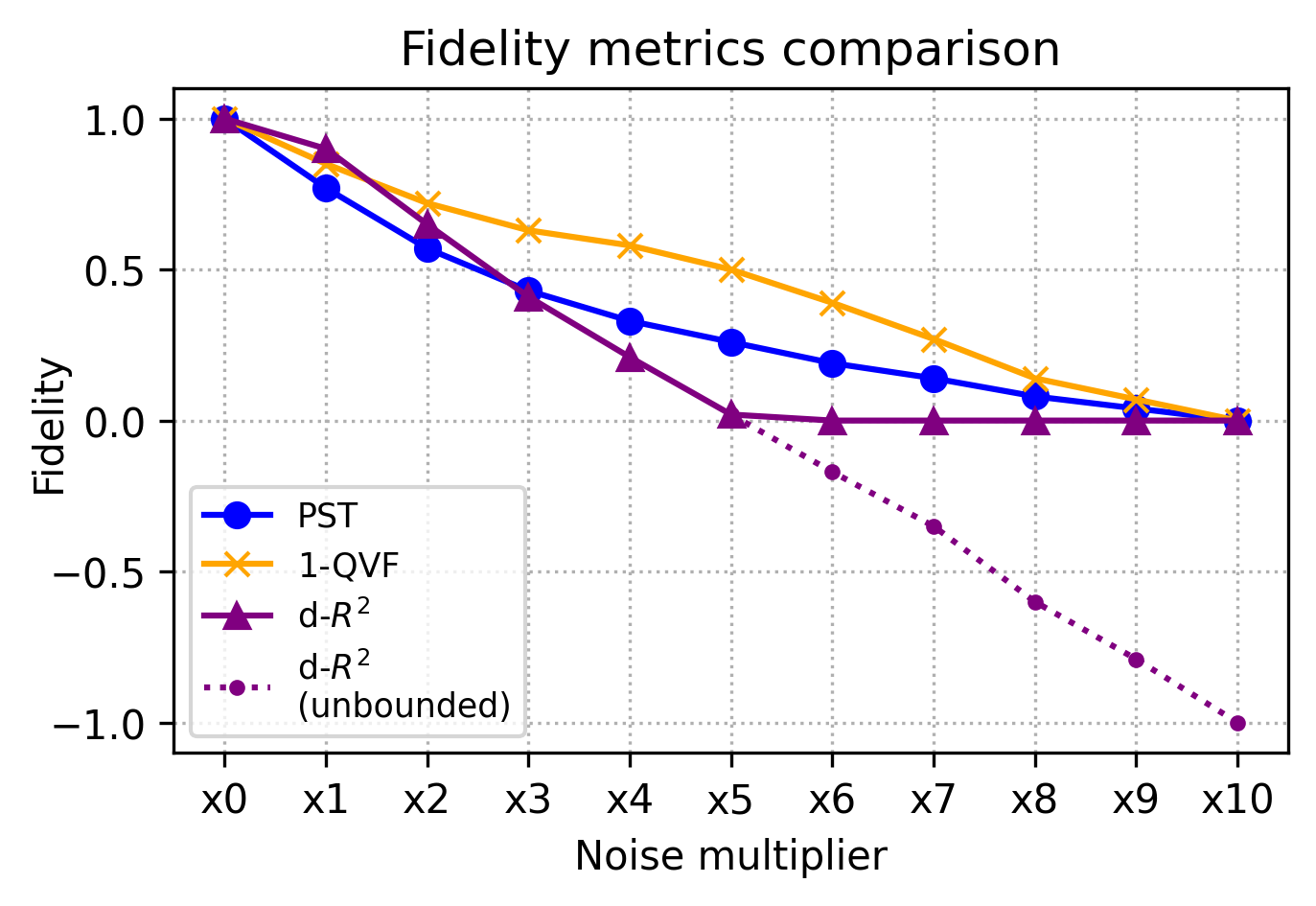}
    \caption{Fidelity metrics comparison when different noise multiplier values are applied on the circuit in Fig. \ref{1a_BV}. Note that d-$R^{2}$ can go to the negative region if the bounding condition in equation \ref{eq_ss} is removed.}
    \label{3d_neg}
\end{figure}

Our modified version of $R^2$ is called discrete-$R^2$ (d-$R^2$). Compared with the other metrics, d-$R^2$ has two important features that make it suitable for NISQ fidelity analysis. First, it takes all the states in the distribution into consideration. So in addition to checking how well the correct states are standing out, d-$R^2$ also penalizes wrong states when they should not appear in the output distribution, which makes it work well with algorithms that output multiple correct states. Second, d-$R^2$ is a measurement of closeness to the uniform superposition, which gives the user d-$R^2=0$ as in Fig. \ref{3d_fids}(d). Because the output distribution of noisy QCs can rapidly converge to the uniform distribution \cite{unifrom_dis, uni_dis_bad_1, uni_dis_bad_2, uni_dis_bad_3}, we think a practical fidelity metric should clearly recognize this worst-case distribution for the user.

\begin{table}[tb]
\caption{discrete-$R^2$ values and proposed interpretations}
\begin{adjustbox}{width=1\columnwidth,center}
\label{tab_r2}
\centering
\begin{NiceTabular}{@{}r|l@{}}
\toprule
d-$R^{2}$				& Interpretation\\ \midrule
$=1$					& output is perfect, same as if there was no noise.\\
$>0.7$ to $\leq 1$		& output is good, the circuit has high fidelity.\\
$>0.5$ to $\leq 0.7$	& output quality is fair, contains noticeable noise.\\
$>0.3$ to $\leq 0.5$	& output contains significant noise, interpret with caution.\\
$>0$ to $\leq 0.3$		& output is extremely noisy, do not use.\\
$=0$					& output is no better than a uniform superposition.\\ \bottomrule
\end{NiceTabular}
\end{adjustbox}
\end{table}

A uniform superposition as in Fig. \ref{3d_fids}(d) does not give any useful information about which state(s) should be the correct state. However, PST and QVF will still show fidelity scores of 0.26 and 0.50. Since every possible output state has equal measurement opportunity, the user can just count the number of qubits in the circuit and randomly pick some states as the output, which invalidates the purpose of executing the quantum circuit in the first place. Sometimes the output distribution fits worse to the expected distribution than a uniform superposition, as in Fig. \ref{3d_fids}(e) and (f). These distributions are caused by defective hardware operations since the error rates for measurements in those QCs are already over 0.5 (0.7 in (e) and 0.9 in (f)). Because such output distribution is faulty and extremely misleading to the user, they should not be used to interpret the QC. We use the bounding condition $SSR<SST$ to limit d-$R^{2}\geq0$ for this practical purpose. However, if the user wants to quantify this faulty distribution, it is still possible to remove the bounding condition and have a negative d-$R^{2}$ score, as shown in Fig. \ref{3d_neg}.

To calculate d-$R^{2}$, we first calculate two \textbf{S}um of \textbf{S}quares ($SS$):

\begin{equation} \label{eq_ss}
\begin{aligned}
SS_{residual}=SSR&=\sum_{i=0}^{2^{n}}\left(Y_{i}-y_{i} \right)^{2}\\
SS_{total}=SST&=\sum_{i=0}^{2^{n}}\left(Y_{i}-mean({Y}) \right)^{2}
\end{aligned}
\end{equation}

Here, $n$ is the number of measured qubit(s), $Y$ is the distribution containing all the measurement counts from the noise-free output, and $\textit{y}$ is the distribution containing all the measurement counts from the noisy output. For example in Fig. \ref{3d_fids}(b), $Y$ would be \((0, 0, 1024, 0)\), so \(mean(Y)\) is \(256\), and $\textit{y}$ would be \((137, 17, 789, 81)\). Note that when calculating the subtractions, the indices of $Y$ and $\textit{y}$ must be aligned so that the measured states match each other. Then d-$R^{2}$ can be obtained as:

\begin{equation} \label{eq_r2}
\begin{aligned}
\text{d-}R^2=\begin{cases}
    1-\frac{SSR}{SST}, & \text{if}\ SSR< SST\\
    0, & \text{otherwise}
\end{cases}
\end{aligned}
\end{equation}

Compared with other metrics empirically, the fixed definition of d-$R^2=0$ as the uniform superposition gives finer granularity for calculations and allows the fidelity interpretations to stay invariant when applied to different circuits. Based on the observations during our experiments and the analysis in Fig. \ref{3d_fids}, we offer the recommended interpretations for different d-$R^{2}$ values used in this paper in Table \ref{tab_r2}.

When the QC is designed to output a uniform superposition, like the Quantum Random Number Generator (QRNG) \cite{QRNG}, the output with no noise and the output with extreme noise can be the same (both are uniform superposition), since heavy noise will cause the output distribution to eventually converge into the uniform distribution \cite{unifrom_dis, uni_dis_bad_1, uni_dis_bad_2, uni_dis_bad_3}. In this case, the output is both perfectly correct and incorrect at the same time, so the definition of fidelity contradicts itself, causing d-$R^{2}$ to become undefined by design due to $SST=0$. 

\begin{algorithm}[tb]
\caption{Q-fid Training Workflow}\label{alg1}
\begin{algorithmic}[1]
    \footnotesize
    \Statex \textbf{Q-fid Training}
    \State \textbf{Input:} quantum circuit dataset  \\\Comment{circuits and corresponding noisy output}
    \State \textbf{Output:} trained Q-fid system
    
    \Statex \textbf{Step 1: Input Preparation}
    \For{each QC in dataset}
        \State noisy\_fidelity(d-$R^{2}$) $\gets$ ideal noise-free output  \\\Comment{compare noisy/noise-free output to calculate training label}
        \State compressed\_QC $\gets$ compress(QC)  \\\Comment{combine parallel gates to the same circuit layer}
        \State text\_QC $\gets$ gate\_to\_text(compressed\_QC)  \\\Comment{map gates to corresponding text label}
    \EndFor

    \Statex \textbf{Step 2: Tokenizer Training}
    \State tokenizer $\gets$ initialize\_tokenizer()
    \For{each text\_QC in dataset}
        \State trained\_tokenizer $\gets$ tokenizer.fit\_on\_texts(text\_QC)  \\\Comment{train tokenizer on all circuits to generate dictionary}
    \EndFor

    \Statex \textbf{Step 3: Tokenization}
    \State tokenized\_QCs $\gets$ []
    \For{each text\_QC in dataset}
        \State tokenized\_QC $\gets$ trained\_tokenizer.texts\_to\_int(text\_QC)
        \State tokenized\_QCs.append(tokenized\_QC)  \\\Comment{translate QC to integer representation}
    \EndFor

    \Statex \textbf{Step 4: Q-fid Training}
    \For{each tokenized\_QC in tokenized\_QCs}
        \State embedded\_QC $\gets$ embedding(tokenized\_QC)  \\\Comment{apply embeddings to integer labels}
        \State lstm\_output $\gets$ LSTM(embedded\_QC.concat())  \\\Comment{process through LSTM}
        \State fidelity\_prediction $\gets$ fully\_connected(lstm\_output)  \\\Comment{final fidelity prediction}
        \State model.fit(noisy\_fidelity) $\gets$ fidelity\_prediction  \\\Comment{fit model against noisy output fidelity score}
    \EndFor

    \State \textbf{Return} trained\_Q\_fid
\end{algorithmic}
\end{algorithm}

\begin{algorithm}[tb]
\caption{Q-fid Inference Workflow}\label{alg2}
\begin{algorithmic}[1]
    \footnotesize
    \Statex \textbf{Q-fid Inference}
    \State \textbf{Input:} Quantum circuits (QCs)
    \State \textbf{Output:} Fidelity predictions of QC

    \Statex \textbf{Step 1: Input Preparation}
    \For{each QC in QCs}
        \State compressed\_QC $\gets$ compress(QC)
        \State text\_QC $\gets$ gate\_to\_text(compressed\_QC)
    \EndFor

    \Statex \textbf{Step 2: Tokenization}
    \State tokenized\_QCs $\gets$ []
    \For{each text\_QC in text\_QCs}
        \State tokenized\_QC $\gets$ trained\_tokenizer.texts\_to\_int(text\_QC)
        \State tokenized\_QCs.append(tokenized\_QC)
    \EndFor

    \Statex \textbf{Step 3: Q-fid Processing}
    \For{each tokenized\_QC in tokenized\_QCs}
        \State embedded\_QC $\gets$ embedding(tokenized\_QC)
        \State lstm\_output $\gets$ LSTM(embedded\_QC.concat())
        \State fidelity\_prediction $\gets$ fully\_connected(lstm\_output)
    \EndFor

    \State \textbf{Return} fidelity\_prediction
\end{algorithmic}
\end{algorithm}

\subsection{Text-based Representation of Quantum Circuits}\label{qcts}

We represent QC in text string formats that are suitable for feeding into the LSTM networks. The QC is first compressed to its true depth, because qubits that are not physically connected in hardware can execute gates in parallel. Then, Q-fid uses a simple and effective protocol to label the gates inside a QC: a short string describing the gate function followed by the qubit index describing where the gate is executed. For example, a Hadamard gate placed on qubit 2 is labeled as \texttt{h2}, a CNOT gate with control on qubit 0 and target on qubit 3 will be \texttt{cx03}. This representation also reflects the layout of the device. For \textit{ibm\textunderscore nairobi}, there will never be a label called \texttt{cx34}, because this connection does not exist in the hardware. This helps the LSTM network to learn the processor’s layout implicitly. Our text format is very similar to many existing quantum programming languages including openQASM and Qiskit, so preparing the text representation of an existing QC for Q-fid is straightforward:

\begin{enumerate}
\item A tokenizer reviews the full QC dataset and builds a dictionary that maps each text label of a gate to a unique integer. The labels are ranked according to the frequency they appear in the circuit, with the most common appearing label mapped to 1. A special label in Q-fid is \texttt{none}, which means the qubit is staying idle. This label is important because the qubit can decay and become noisier even when no gates are applied. An example of this workflow is shown in Fig. \ref{3a_flow} from (a) to (f).
\item For a large QC dataset, the depth of the individual QCs can vary a lot. To improve the accuracy and efficiency of LSTM training, it is better to fix the number of time steps in the dataset. Once the maximum number of time steps is set, all of the integer-based QC vectors are either truncated or padded to the same depth. The reason for the tokenizer to start from 1 is that 0 is reserved as the padding element, so that the LSTM can safely skip it. We follow the common practices in NLP \cite{padding} to use pre-padding and post-truncation.
\item After all the text labels are converted to integers, every QC in the dataset is now a dense vector. However, the integers do not possess any relationship or similarity information between different gates, so it is hard for the LSTM to learn the effect of interconnected quantum gates. This problem is solved by using word embeddings, a technique of using a higher dimension fixed-length vector to replace the integer encoding \cite{word2vec}. The values of this embedding vector are trained with the LSTM in parallel, which helps Q-fid capture more detailed information from the QC. The embedding layers are shown in Fig. \ref{3a_flow}(g).
\end{enumerate}

This tokenization process can represent a QC in human-readable text format and effectively retains vital circuit structural information, therefore making it possible for neural networks designed for time-series data to process a QC directly. More importantly, it enables existing NLP neural networks to perform analytical tasks on QCs without complex modification. The training of the tokenizer is expressed in Algorithm \ref{alg1}, and the inference process is described in Algorithm \ref{alg2}, all as part of the full Q-fid workflow.

\begin{figure}[tb]
    \centering
    \includegraphics[width=1\columnwidth]{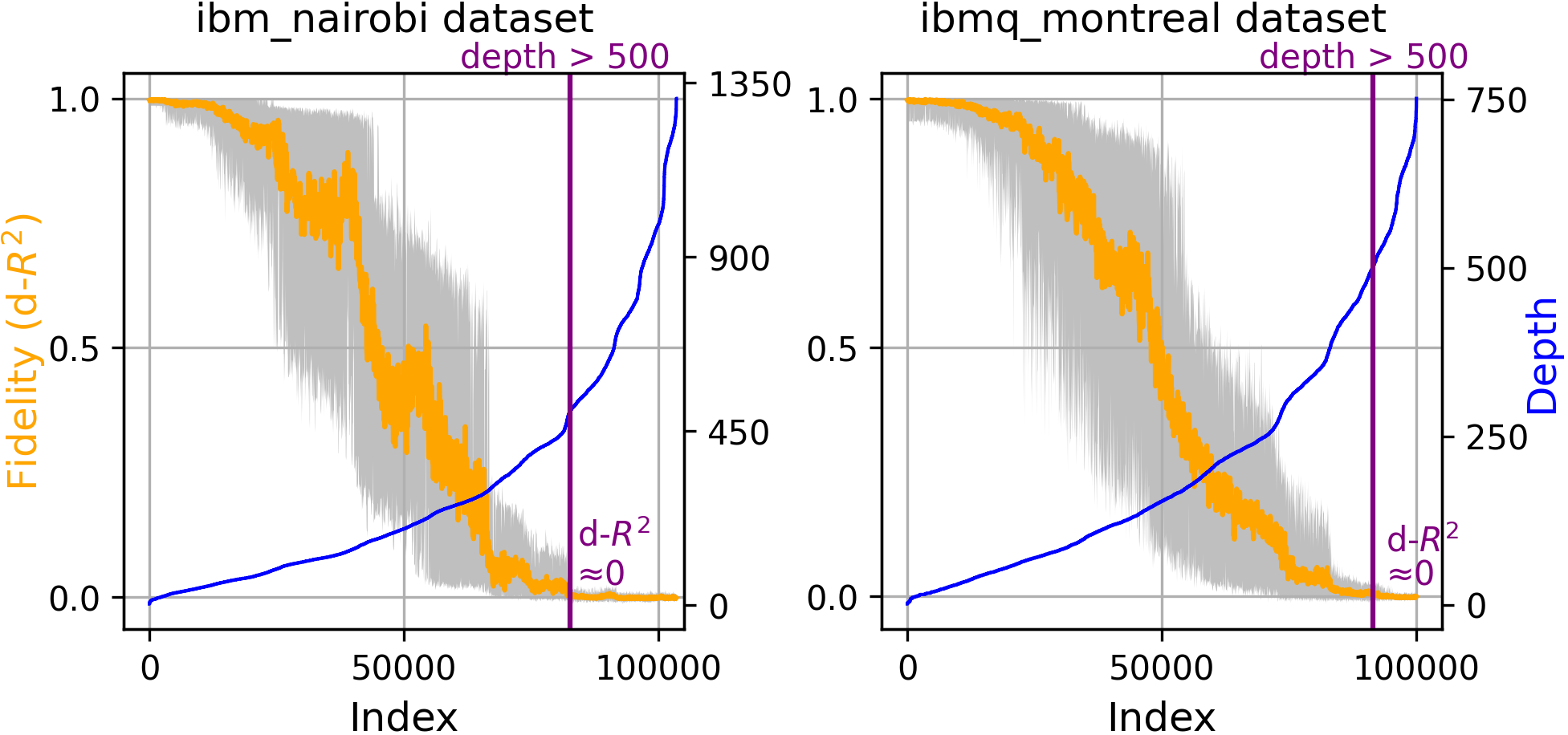}
    \caption{Two QC datasets showing the fidelity of a QC decreases with increasing depth. The x-axis represents the indices of RB circuits ordered from the shallowest to the deepest ones, with a vertical purple line indicating the cut-off point of \(depth = 500\). The fidelity has a very large variance so the raw data is plotted using a gray envelope and an orange average line.}
    \label{4a_dataset}
\end{figure}

\subsection{Training Circuits and Dataset for Q-fid}

Training Q-fid requires a large QC dataset that is also diversified in circuit depth and width, as the relationship between fidelity and circuit size is not necessarily linear. In this work, we utilize the RB circuits to efficiently create a QC dataset to train Q-fid. Because the gates in RB circuits are randomly generated, they span over all different types of gates, depths, widths, and qubit interactions. This provides the required diversity of width and depth for the dataset. In our generated \textit{ibm\textunderscore nairobi} dataset, all the available gates [X, RZ, SX, CNOT, Measurement] are sufficiently covered on all possible qubits, resulting in 40 total unique text labels plus the placeholder \texttt{none} label. The most common gate label in the dataset is \texttt{rz3} with \(2,341,798\) appearances, and the least common label is \texttt{m4} with \(29,528\) appearances.

Using RB circuits in training provides many benefits, one of the advantages is that it greatly simplifies fidelity calculation. Every RB circuit is by definition equivalent to an identity operator, so if the qubits are initialized to $\ket{0...0}$, we know the ideal output states must also be $\ket{0...0}$. This means that in order to calculate the ground-truth training labels (d-$R^{2}$ scores) for an RB circuit, we only need to obtain one noisy output distribution of the circuit. In addition, most of the computing complexity for generating RB circuits comes from calculating the final circuit block that reverses the previous operations, and it can be calculated efficiently in polynomial time, proved by the Gottesman-Knill theorem \cite{GK}.


\section{Methodology}

\begin{figure}[tb]
    \centering
    \includegraphics[width=1\columnwidth]{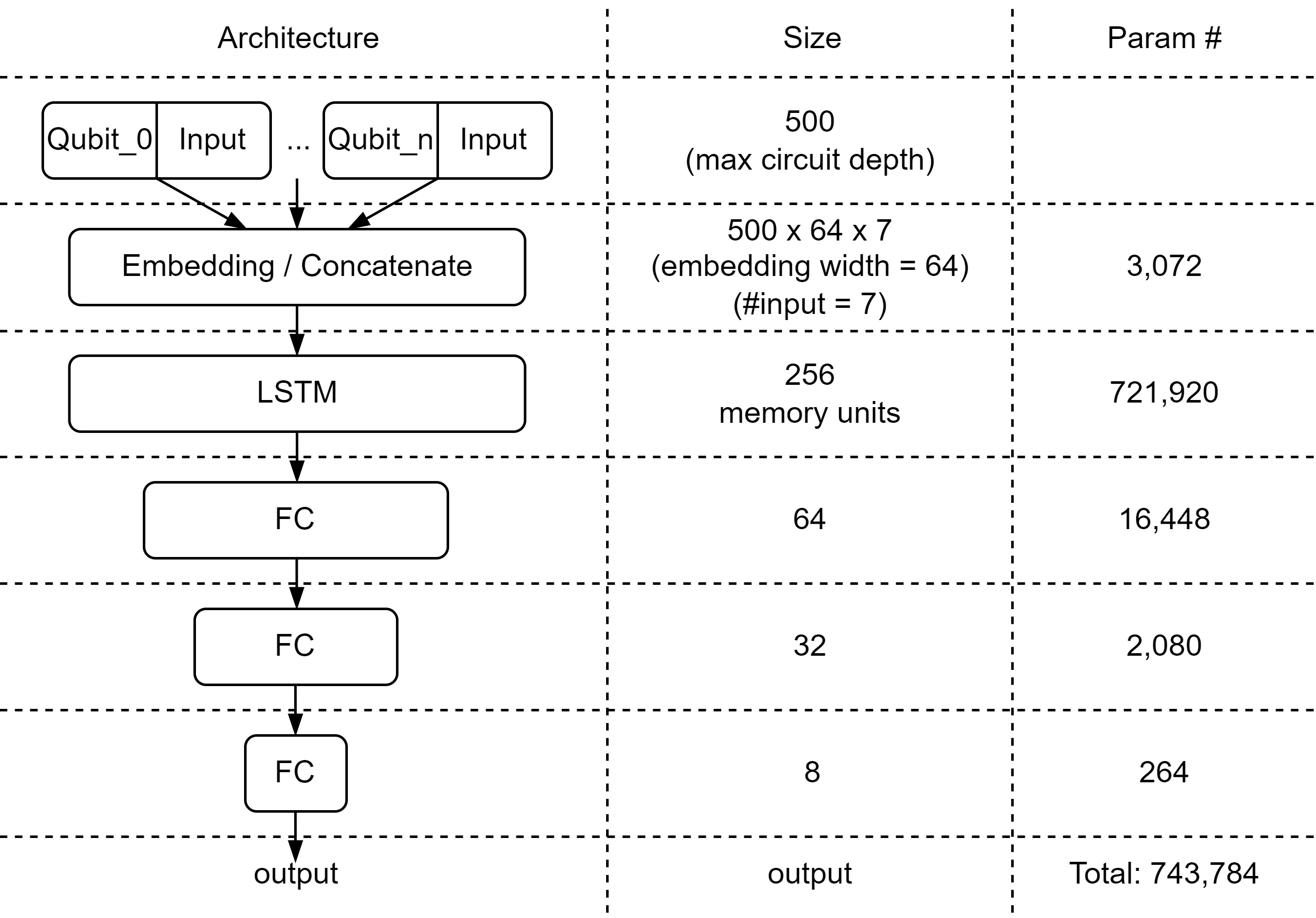}
    \caption{Neural network architecture of Q-fid. FC stands for Fully-Connected layer. The number of input layers depends on the number of qubits inside a processor. This figure shows the architecture for {\textit{ibm\textunderscore nairobi}} with 7 input neurons since it is a 7-qubit processor. The hardware qubit layout is shown in Fig. S4.}
    \label{4b_arch}
\end{figure}

\subsection{Dataset} \label{dataset}

We created two datasets based on two real IBM Quantum processors, \textit{ibm\textunderscore nairobi} and \textit{ibmq\textunderscore montreal}. The QCs are randomly generated according to the RB protocol, with the length of the RB sequence ranging from 1 to 5. We also change the number of active qubits when generating the RB circuits. For example, even though \textit{ibm\textunderscore nairobi} is a 7 qubit processor, we generated RB circuits that only require 1 qubit, 2 qubits, etc. The main reason for doing this is because placing the QC on different physical qubits on the same processor can yield different fidelity due to noise and manufacture variation. When the RB circuit only requires 1 active qubit, we can place the circuit on 7 different physical qubits on \textit{ibm\textunderscore nairobi}. This improves Q-fid's ability to learn the properties of individual qubits.

Due to the ever-changing nature of quantum errors, instead of depending on a fixed noise model to predict the fidelity of a quantum circuit, we designed Q-fid to actively learn the noisy output results from a quantum processor and incorporate the gate/qubit error parameters inside its hidden layers. During training, Q-fid needs the fidelity (d-$R^2$) score as the training label corresponding to the input circuit. This fidelity score is calculated by comparing the noisy output result to the theoretical noise-free result, which means that if the noise characteristics for a particular processor have changed, Q-fid will learn this change from the updated fidelity score and give new predictions based on the latest processor characteristics.

The \textit{ibm\textunderscore nairobi} QC dataset contains 103,500 circuits, and the \textit{ibmq\textunderscore montreal} QC dataset contains 100,000 circuits. All of the circuits are transpiled and converted to a text-based representation described in section \ref{qcts}, then the noisy output results are captured using the Qiskit Aer simulator. We constrain the depth of the circuits to be less than or equal to 500 because circuits deeper than 500 have near-zero fidelity, as shown in Fig \ref{4a_dataset}. After trimming, the \textit{ibm\textunderscore nairobi} QC dataset has 82,644 circuits, and the \textit{ibmq\textunderscore montreal} QC dataset has 91,386 circuits.

\begin{figure}[tb]
    \centering
    \includegraphics[width=1\columnwidth]{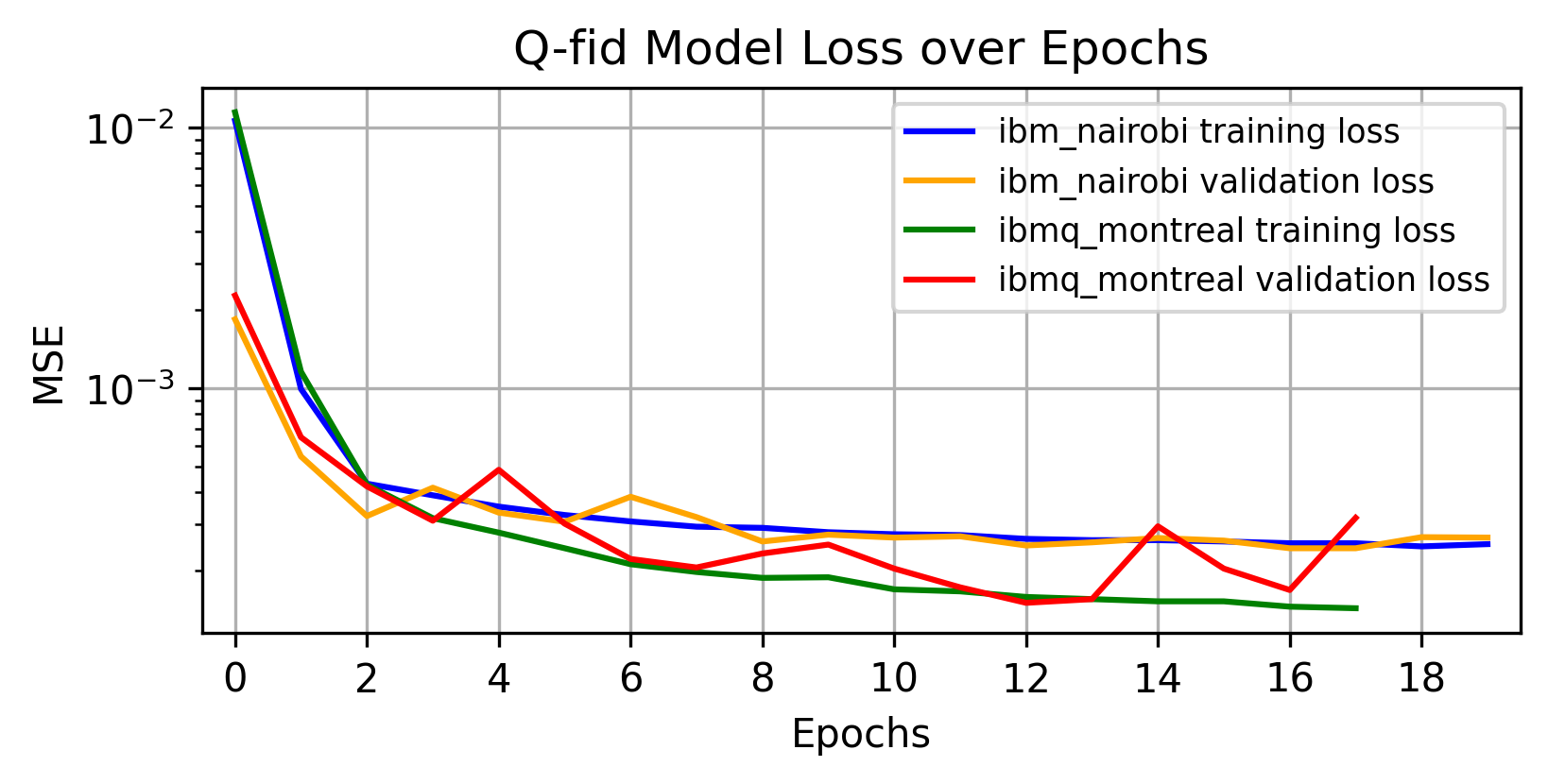}
    \caption{Training loss and validation loss for the two datasets. Training terminated on epoch 20 for \textit{ibm\textunderscore nairobi}, and on epoch 18 for \textit{ibmq\textunderscore montreal}. The final validation loss  is \(0.150\)E-3 for \textit{ibm\textunderscore nairobi}, and \(0.243\)E-3 for \textit{ibmq\textunderscore montreal}}
    \label{4b_loss}
\end{figure}

\begin{figure}[tb]
    \centering
    \includegraphics[width=1\columnwidth]{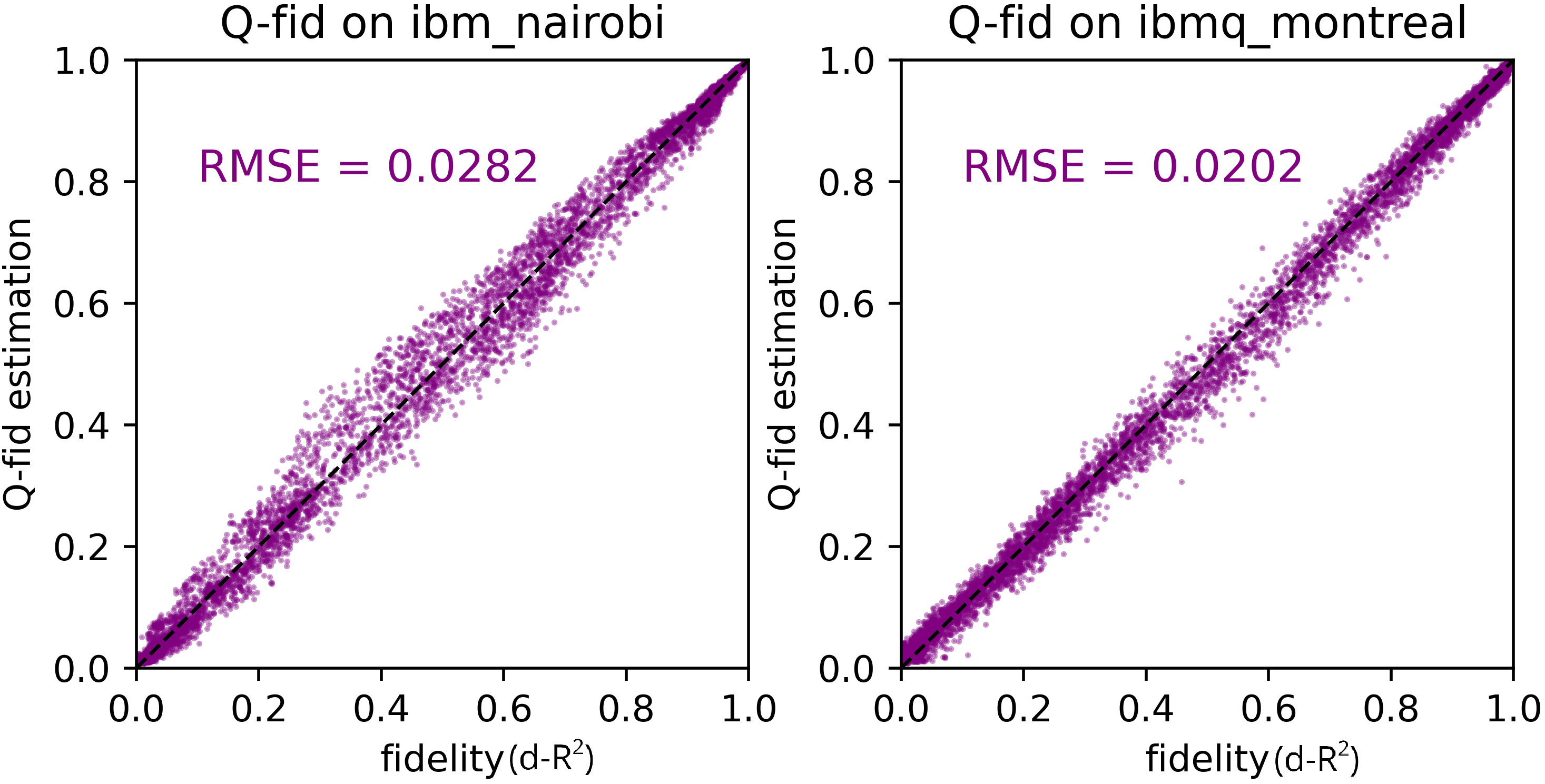}
    \caption{Scatter plot of real fidelity vs. predicted fidelity of Q-fid running on the test set. The test set for \textit{ibm\textunderscore nairobi} contains 8,265 circuits, and the test set for \textit{ibmq\textunderscore montreal} contains 9,140 circuits}
    \label{4b_testset}
\end{figure}

\subsection{Model and Training}\label{training}

We build two Q-fid models to test the two processors, the architecture of Q-fid for \textit{ibm\textunderscore nairobi} is shown in Fig. \ref{4b_arch}. Each qubit has its own input neuron and they all have a length of 500, equal to the maximum QC depth of the dataset. The embedding layer transforms every gate label into a 64-dimensional dense vector and they are all concatenated together to represent one input timestep. The LSTM layer uses 256 memory units, followed by three Fully-Connected layers. We use ReLU activation and Sigmoid activation for the final layer. The model for \textit{ibm\textunderscore nairobi} contains 743,784 trainable parameters with 7 input layers, and the model for \textit{ibmq\textunderscore montreal} contains 7,403,004 trainable parameters with 27 input layers. The model for \textit{ibmq\textunderscore montreal} has mostly the same architecture, only changing the number of input layers and adding the number of hidden units.

To demonstrate that no special modification is needed for existing LSTM architectures to analyze a quantum circuit, we picked the original implementation of LSTM with their standard forget/input/output gates by Hochreiter and Schmidhuber \cite{lstm}, we also implemented the text tokenizer and the embedding layer using the TensorFlow Keras API. The models run on a rack server with two Intel Xeon Gold 6354 processors and the Nvidia A100 GPU. The training workflow can be reproduced using the Jupyter Notebook available online mentioned in Sec. \ref{code}. The models are trained with a batch size of 32 and Adam optimizer for 20 epochs using the MSE loss function, the training automatically terminates if the loss does not improve for 5 continuous epochs. The training, validation, and test split ratio is 7:2:1, we evaluate loss on the validation set and pick the best-performing model on the test set. On average, each epoch took 81s to train and the inference time per circuit was 0.48ms. The training curve for the two models is shown in Fig. \ref{4b_loss}, and Fig. \ref{4b_testset} shows the scatter plot of real fidelity vs. predicted fidelity of Q-fid running on the test set.


\section{Results \& Discussion}\label{demos}

\begin{figure}[tb]
    \centering
    \includegraphics[width=\columnwidth]{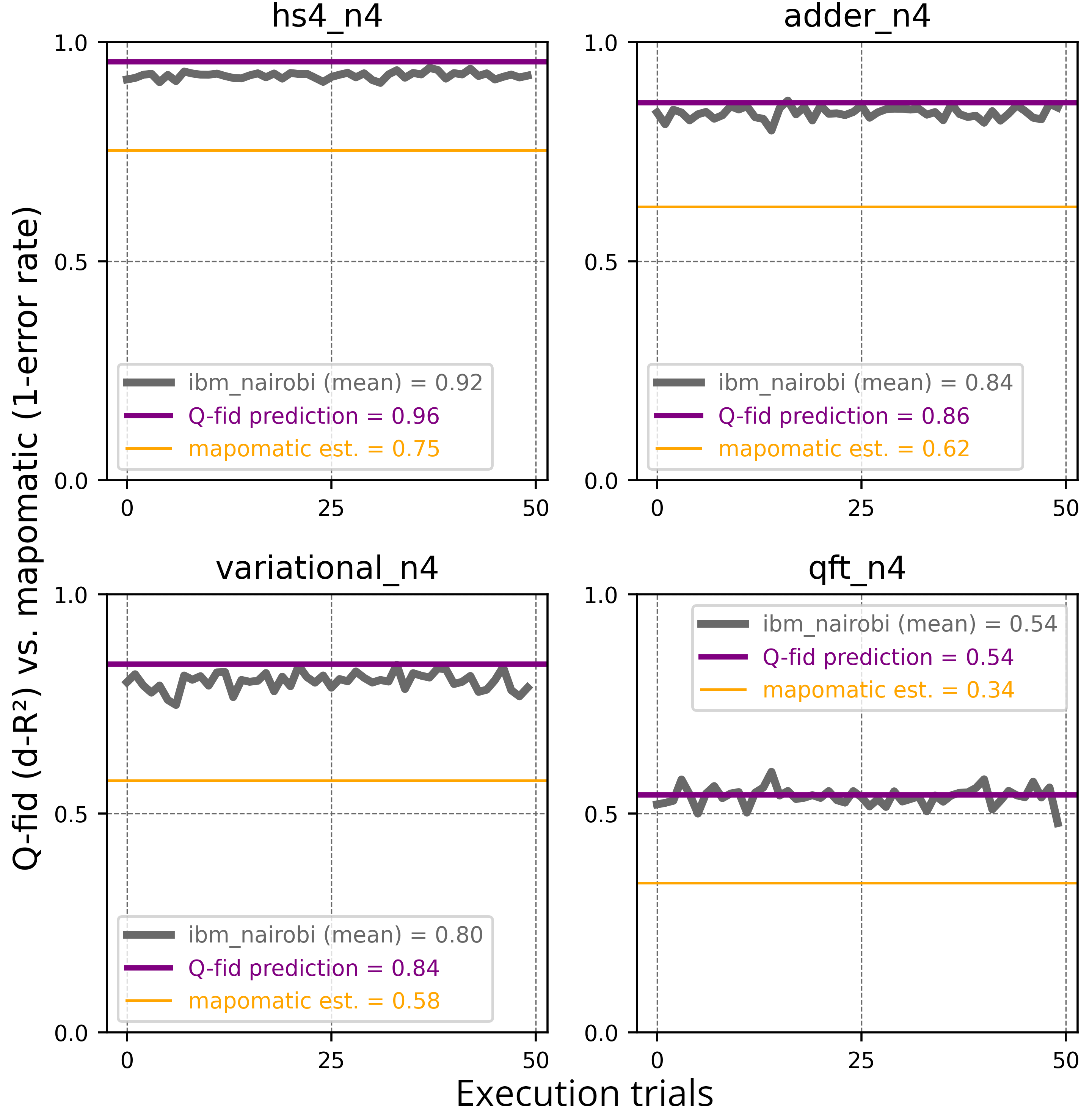}
    \caption{Q-fid's fidelity prediction (purple) compared with \texttt{mapomatic} (orange) on \textit{ibm\_nairobi}. Due to the probabilistic nature of quantum noise, the fidelity for the same circuit oscillates when executed with the same device noise model (gray). Each circuit is executed with 1,024 shots to produce a noisy output distribution, then the mean fidelity is calculated from 50 noisy distributions (x-axis) as d-$R^{2}$. Q-fid accurately predicts the mean fidelity of each circuit.}
    \label{5a_fid}
\end{figure}

Based on the number of qubits, connection complexity, and algorithm practicality, we picked 25 quantum circuits from the \texttt{QASMbench} \cite{qasmbench} NISQ benchmark suite to demonstrate the performance of Q-fid. We compare our result with \texttt{mapomatic}, the latest and default mapping tool of the Qiksit transpiler. \texttt{mapomatic} predicts circuit fidelity and maps high-fidelity circuits onto the quantum processor, the prediction is made by accumulating individual gate error rates of the circuit \cite{mapgit}, which has the same prediction range and statistical meaning as d-$R^{2}$: 1 predicts a perfect output distribution equivalent to the noise-free simulation, and 0 predicts a distribution with maximum possible error rates, equivalent to the uniform distribution. The circuits are executed on the Qiskit Aer simulator \cite{qiskit} with a noise model that mimics \textit{ibm\_nairobi} and \textit{ibmq\_montreal}. Section \ref{5a} demonstrates Q-fid's prediction performance on \textit{ibm\_nairobi}. Section \ref{5b} demonstrates Q-fid's ability to correctly find high-fidelity transpilation layouts on \textit{ibmq\_montreal}. Section \ref{5c} demonstrates how Q-fid adapts to \textit{ibm\_nairobi}'s device variance on different dates. The four representative samples picked from the full result are: Hidden Subgroup problem (\texttt{hs4\_n4}) \cite{scafcc}, Quantum Ripple Carry Adder (\texttt{adder\_n4}) \cite{scafcc}, Variational Ansatz (\texttt{variational\_n4}) \cite{openfermion}, and Quantum Fourier Transform (\texttt{qft\_n4}) \cite{openqasm3}. 

\begin{figure}[tb]
    \centering
    \includegraphics[width=\columnwidth]{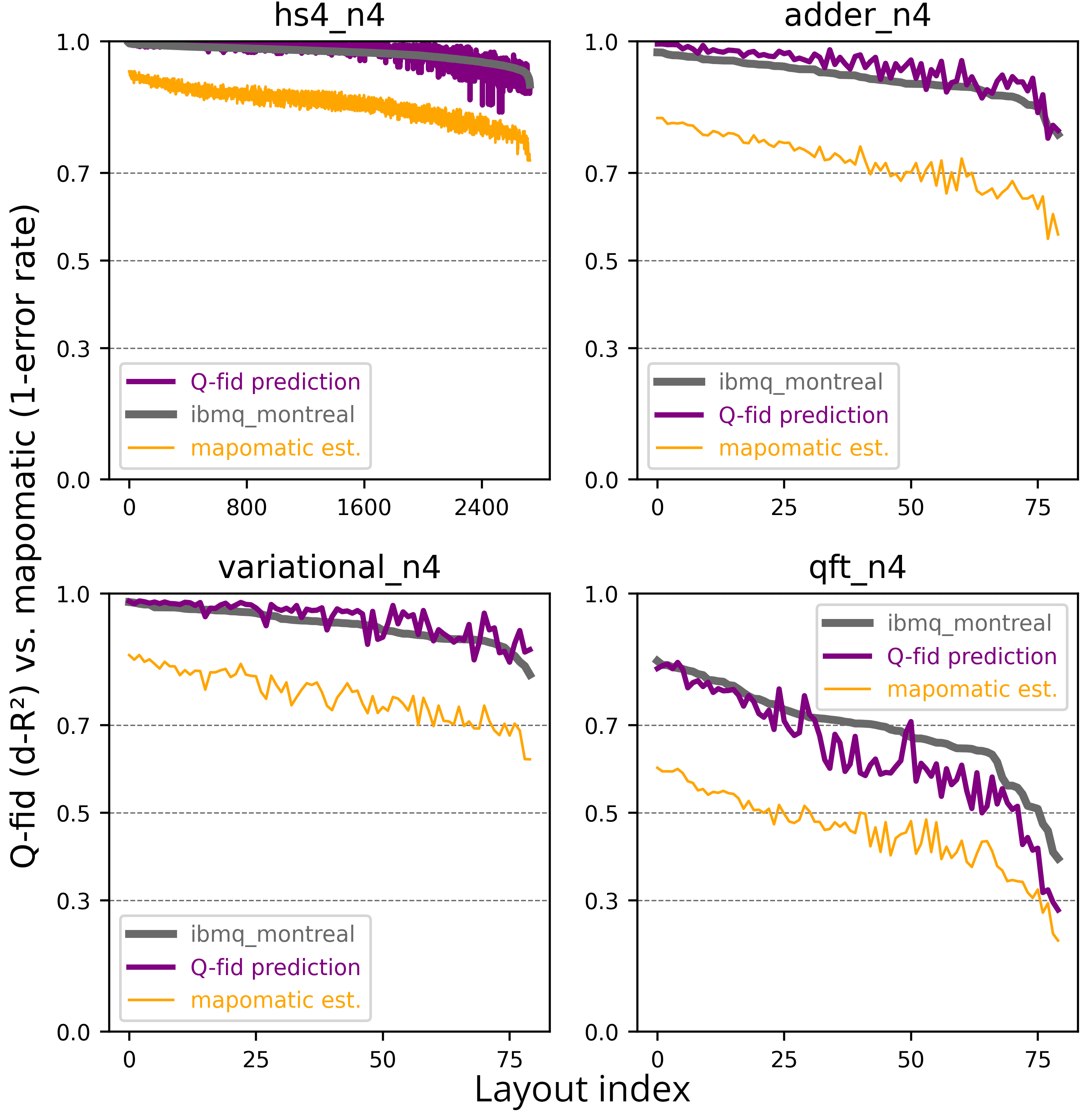}
    \caption{Q-fid's fidelity prediction (purple) for different circuit layouts on \textit{ibmq\_montreal}. The x-axis is the index of layouts for the quantum circuit, the y-axis shows different d-$R^{2}$ fidelity score regions listed in table \ref{tab_r2}. The order of the data is sorted from the highest fidelity layout to the lowest fidelity layout according to the fidelity reported by \textit{ibmq\_montreal} (gray). Q-fid accurately tracks the fidelity variance of every layout for each circuit.}
    \label{5b_layout}
\end{figure}

\subsection{Fidelity Prediction} \label{5a}

Fig. \ref{5a_fid} demonstrates Q-fid's prediction performance on \textit{ibm\_nairobi}. For one given circuit, both Q-fid and \texttt{mapomatic} can only give one static fidelity prediction. Under this circumstance, comparing the performance between Q-fid and \texttt{mapomatic} using the mean hardware execution fidelity results from multiple trials shows the effect of quantum noise and how the fidelity prediction can provide a representative guideline. All the circuits used in this experiment are not optimized for physical qubit layout, so the logical qubits are placed on \textit{ibm\_nairobi} in numerical order from physical qubit 0 to 6 and stay on the same layout for all 50 trials.

Because most of the test circuits are placed on the same physical qubits, the variables in this experiment are the different quantum gates used in different algorithms. The results prove that Q-fid has learned the noise characteristics of different quantum gates on the same physical qubit from the training circuits. In contrast, the error rate based \texttt{mapomatic} tends to give an underestimation, especially for high-fidelity circuits.

\begin{figure}[tb]
    \centering
    \includegraphics[width=\columnwidth]{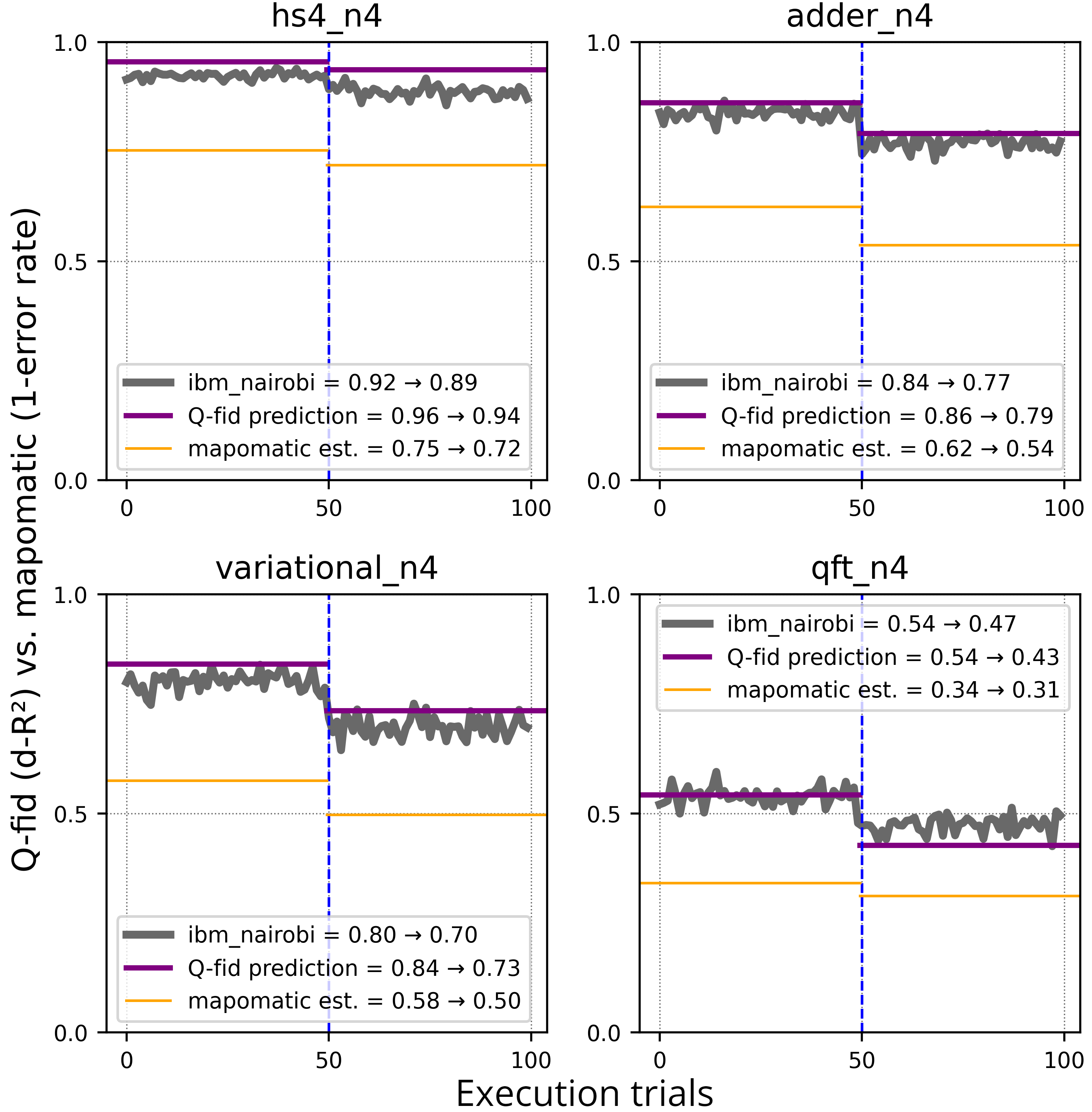}
    \caption{Q-fid's fidelity prediction (purple) compared with \texttt{mapomatic} (orange) on \textit{ibm\_nairobi}. Again, the fidelity oscillates (gray) due to the probabilistic nature of quantum noise. The first 50 trials (left of the blue line) are performed on Nov. 15, 2022, which is the same data as in Fig. \ref{5a_fid}. The next 50 trials (right of the blue line) were performed on Nov. 18, 2022, which shows that \textit{ibm\_nairobi}'s performance is slightly worse due to different noise patterns. Q-fid adapts to this new noise pattern and it only takes 100 new RB circuits to retrain.}
\label{5c_adaptive}
\end{figure}

\subsection{Transpilation Optimizations} \label{5b}

Fig. \ref{5b_layout} shows Q-fid's ability to correctly find high-fidelity transpilation layouts on \textit{ibmq\_montreal}. Since \textit{ibmq\_montreal} is a 27-qubit processor, a circuit only using 4 physical qubits can have many layout options depending on different transpilations. For example, the \texttt{hs4\_n4} circuit has a depth of 34 and 13 CNOT gates, and it has a total of 2,728 different layouts shown on the x-axis, ranked from the highest fidelity layout to the lowest fidelity layout. 
\texttt{mapomatic} relies on the latest hardware calibration data to calculate the fidelity, and uses the relative fidelity differences to find the high-fidelity layouts. In comparison, Q-fid achieves similar performance without explicit hardware calibration data input, and gives better absolute fidelity prediction value so the user has more potentially good layouts to pick from. Additionally, from the full result in Fig. S2, it shows that \texttt{mapomatic}’s prediction is not sensitive to the layout variance of the deep and complex circuits, which makes it a less preferable metric for training neural networks.

In section \ref{5a}, the circuits are executed on fixed physical layouts, so the objective is to test if Q-fid can learn the effects of different quantum gates when they are applied to the same physical qubits. However, in this experiment, the same circuits are executed with different layouts, so the new objective here is to test if Q-fid can learn the noise characteristics when the same quantum gates are placed on different physical qubits. 

\begin{figure*}[htb]
    \centering
    \includegraphics[width=1\textwidth]{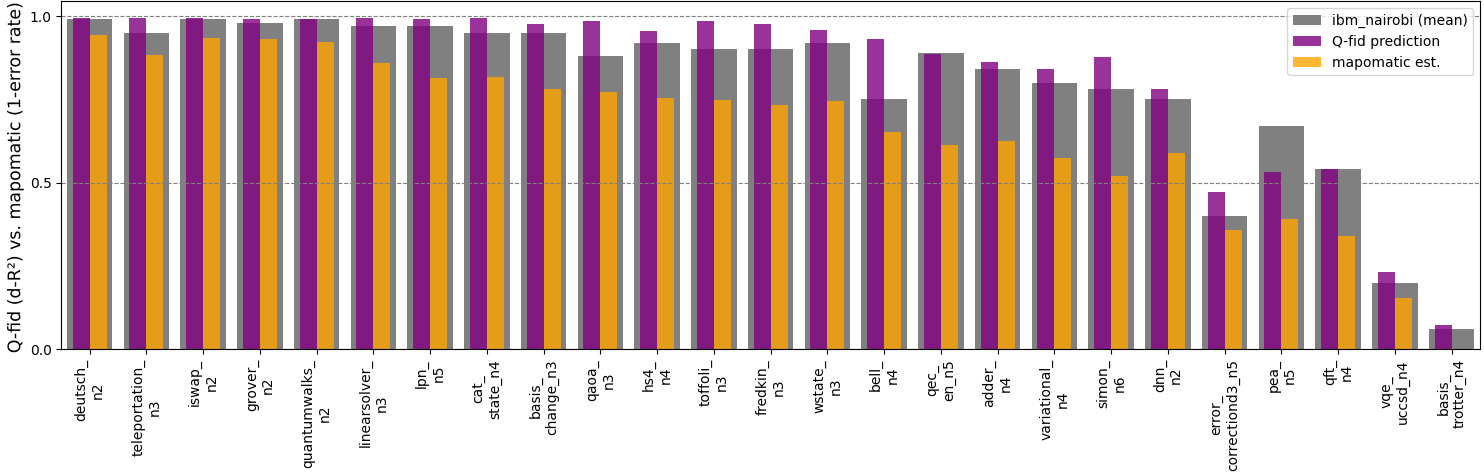}
    \caption{Result comparison between the mean noisy fidelity from \textit{ibm\_nairobi} and the predictions from Q-fid/\texttt{mapomatic}, the mean fidelity for each circuit is calculated from 50 noisy outputs as d-$R^{2}$. The RMSE of Q-fid's prediction compared with the mean noisy fidelity ranges from \(0.003 \) to \(0.182\). On the other hand, \texttt{mapomatic}'s prediction has a minimum RMSE of \(0.0424\) and a maximum RMSE of \(0.284\).}
    \label{5a_result}
\end{figure*}

\subsection{Noise-adaptive Training} \label{5c}

Fig. \ref{5c_adaptive} demonstrates how Q-fid adapts to \textit{ibm\_nairobi}'s device variance on different dates. The same experiment in section \ref{5a} is performed again, but the noise model of \textit{ibm\_nairobi} is changed to Nov. 18, 2022, three days later than the first set of predictions. The new noise model has a slightly worse fidelity performance than the previous one, which is shown to the right of the blue vertical line. For the new noise model, Q-fid's prediction is \(5.31\times\) more accurate than \texttt{mapomatic} on average, with the most accurate one being \(42.0\times\) better.

We envision that Q-fid can be constantly learning in parallel with the execution of QCs inside a quantum processor, so that it can always give predictions based on the most recent hardware characteristics. For this experiment, 100 additional RB circuits are randomly generated according to section \ref{dataset}, then executed on \textit{ibm\_nairobi} with the new noise model. These new execution data go under the same training process as described in section \ref{training} to retrain Q-fid. For different processors, the number of circuits needed for retraining might vary. However, because Q-fid can use the output of any historical workload to update its internal parameters, it can be updated directly and continuously in parallel with the processor's normal workflow. In contrast, \texttt{mapomatic} needs to use the latest gate error rates to give updated predictions, which requires the processor to perform individual RB experiments for single-qubit gates, two-qubit gates, and the subsequent data-fitting calculations \cite{rb_0}. This process interrupts normal workflow and the prediction accuracy entirely depends on how recently the calibration jobs are performed. 

\subsection{Summary}

Our comparison shows that although \texttt{mapomatic} is doing well in following the trend of fidelity degradation due to heavy noise, it gives an underestimation which causes the user to mistakenly think the circuit is too noisy to execute, but the output result will be relatively acceptable. An overview of the result comparison between the mean noisy fidelity and the predictions from Q-fid/\texttt{mapomatic} is shown in Fig. \ref{5a_result}. The RMSE of Q-fid's prediction compared with the mean noisy fidelity ranges from \(0.003 \) to \(0.182\) with an average of \(0.0515\). On the other hand, \texttt{mapomatic}'s fidelity estimation has an average RMSE of \(0.142\), with a minimum RMSE equal to \(0.0424\) and a maximum RMSE of \(0.284\). On the Quantum Walks algorithm (\texttt{quantumwalks\_n2}) \cite{raffmiceli}, Q-fid's prediction is \(24.7\times\) more accurate than \texttt{mapomatic}. On \textit{ibmq\_montreal}, when finding the high-fidelity layouts, Q-fid correctly finds the top 10\% of high-fidelity circuit layouts for all 25 algorithms. Within those top 10\% of layouts, Q-fid's predicted fidelity has an average RMSE of \(0.0252\), up to \(32.8\times\) more accurate than \texttt{mapomatic}.


\section{Related Work}

Randomized Benchmarking is one of the earliest experiments developed to characterize quantum operation error rates \cite{rb_2, rb_1, rb_0}, and remains the most commonly used tool for this purpose today. The latest works focus on improving Randomized Benchmarking to support more quantum operations \cite{rb_4}, and make the experiment more flexible for larger circuits consisting of many qubits \cite{rb_3, rb_5}.

A QC must be carefully implemented on real hardware to retrieve useful measurements. Early research concentrated on circuit compilation techniques to improve the fidelity of the output. For example, using gate scheduling to reduce the number of physical quantum gates or rerouting the CNOT connections to minimize SWAP gates \cite{error_mitigation_0, error_mitigation_1, error_mitigation_2, error_mitigation_3}. Recently, the research direction has shifted to hardware-specific optimization and noise-aware qubit layout \cite{allocation, error_mitigation_4, error_mitigation_5, error_mitigation_6}. Due to the constantly changing device calibration data, quantum circuits have to be transpiled according to the architecture and noise characteristics of the processor in order to achieve the best performance.

Fidelity estimation is an emerging research field in quantum computing. Although it has been proved that estimating the final fidelity from the QC itself is hard in general, many works have demonstrated using quantum algorithms to attack this problem and achieving exponential speedup \cite{alg_fid_est_0, alg_fid_est_1}. Other approaches like statistical estimation and polynomial fitting were also investigated \cite{fid_est_2, fid_est_3}. 


Using machine learning for fidelity estimation is still a very new area of research. \cite{fid_est_3} proposed a shallow neural network to directly estimate the fidelity of the quantum states, although the quantum states are not prepared with a QC. For our work, we focus on the holistic view of quantum states in circuit model quantum computing. In \cite{onthe}, the viability of using machine learning to predict the state fidelity from circuit representation was proved. This work utilizes a Convolutional Neural Network (CNN) for feature extraction and models the input circuit using integer encoding. The CNN architecture is heavier than our LSTM architecture, and the circuit-to-integer encoding also requires one extra step to implement. In Q-fid, this encoding is automatically handled by the text tokenizer. \cite{usingml} also uses a CNN for feature extraction, but the number of parameters for a 3$\times$3 circuit is almost 12$\times$ more compared with Q-fid's LSTM architecture, rendering future scalability issues. Recent works are exploring different learning methods. \cite{Multimodal} proposes a multimodel deep learning method where they utilize two neural networks to learn from the measurement modality and circuit-encoded modality separately. In Q-fid, the QC is explicitly fed into the neural network and the measurement information is implicitly learned by the network through the corresponding d-$R^2$ score. Another work \cite{fid_est_6} implements the graph transformer for the same fidelity prediction task, but since the input QC is modeled like an image, a deep QC will produce a very long rectangle image and the performance needs to be investigated.


\section{Conclusion \& Outlook}

We present the Q-fid system to accurately predict the fidelity of a quantum circuit running on a real NISQ processor. We show that the performance of NISQ processors is easily affected by external noise, so with Q-fid's fidelity prediction we can help save quantum computing resources by optimizing circuit layout and reducing execution shots. Q-fid uses LSTM to learn the noise properties of the qubit and the relationship between quantum gates, without the need for any separate input of hardware calibration data and gate error rates. A novel method to model the quantum circuits using text labels was presented, and the full training workflow was introduced. We apply the d-$R^2$ metric to intuitively quantify the fidelity of a noisy quantum circuit. Based on this metric, we also showed how to generate a training circuit dataset using the Randomized Benchmarking circuits. We compare Q-fid's performance with \texttt{mapomatic}, and the results prove that Q-fid can effectively learn the characteristics of different qubits, gates, and the structure of quantum circuits. 

Future improvements of Q-fid can focus on the neural network component. The structure of the neural networks can be adjusted and optimized to use fewer layers of parameters, and different tokenizer configurations can be investigated to see how they affect Q-fid’s prediction accuracy \cite{token0, token1}. Because the QC is treated as text inputs in Q-fid, various new LSTM implementations or Recurrent Neural Network (RNN) architectures can also replace the standard LSTM used in this work to improve performance. The goal of Q-fid is to help enhance the usability of current NISQ devices, where the number of qubits is limited and requires classical optimization to overcome noise problems. In the future when fault-tolerant qubits are realized, we will likely need more sophisticated metrics to evaluate the fidelity of QC outputs, due to the exponentially growing problem space accompanied by the increase of qubit counts.


\section{Data and Code Availability}\label{code}

The full results and raw data from section \ref{demos} are available in Supplementary Material: Fig. S1, S2, S3 and Table S1, S2. The layout of the computing devices and their characteristics at the time of the experiments is available in Fig. S4 and Table S3, S4, S5. The datasets used for training Q-fid are available at \url{https://www.kaggle.com/datasets/ykmaoykmao/q-fid-datasets}. The code for generating the datasets is available at \url{https://github.com/yikaimao/Q_fid}.


\section*{Acknowledgments}

This work was supported by JST COI-NEXT (JPMJPF2221), JST SPRING (JPMJSP2123), and JSPS KAKENHI (JP24K20843). We acknowledge the use of IBM Quantum services for this work. The views expressed are those of the authors, and do not reflect the official policy or position of IBM or the IBM Quantum team.



%



\ifCLASSOPTIONcaptionsoff
  \newpage
\fi



\bibliography{refs.bib}{}
\bibliographystyle{IEEEtran}
\end{document}